\documentclass[preprint,12pt]{elsarticle}
\usepackage{tabularx}
\usepackage{multirow}
\usepackage{amssymb}
\usepackage{amsmath}
\usepackage{float}
\usepackage{bm}
\usepackage[left=1.5cm, right=1.5cm, top=2.5cm, bottom=2.5cm]{geometry}
\usepackage{xcolor}
\usepackage{array}
\makeatletter
\def\NAT@sort@citestrue{\let\NAT@sort@cites\@ne}
\def\NAT@cmprs@citestrue{\let\NAT@cmprs@cites\@ne}
\makeatother

\journal{Journal of Alloys and Compounds}

\begin{document}

\begin{frontmatter}

\title{Prediction of Mechanical Properties and Thermodynamic Stability of Ti-N system using MTP Interatomic Potential}

\author[label1,label2]{Pradeep Kumar Rana}
\author[label1,label2]{Atharva Vyawahare}
\author[label1,label2]{Rohit Batra}
\author[label1,label2]{Satyesh K. Yadav}

\affiliation[label1]{organization={Department of Metallurgical and Materials Engineering, Indian Institute of Technology Madras},
            city={Chennai},
            postcode={600036}, 
            state={Tamil Nadu},
            country={India}}

\affiliation[label2]{organization={Center for Atomistic Modeling and Materials Design, Indian Institute of Technology Madras},
            city={Chennai},
            postcode={600036}, 
            state={Tamil Nadu},
            country={India}}
            
\begin{abstract}
Ti-N material system have range of compounds with different stoichiometry like Ti\textsubscript{2}N, Ti\textsubscript{3}N\textsubscript{2}, Ti\textsubscript{6}N\textsubscript{5}, Ti\textsubscript{4}N\textsubscript{3} alongwith Ti , TiN and solid solutions of N in Ti with a maximum of 23\% solubility. In this work, we develop an interatomic potential based on moment tensor potential (MTP) that could reliably predict mechanical properties and thermodynamic stability of all Ti-N system. Taking into account the structural similarity and dissimilarity of various Ti-N system to choose training dataset was crucial for development of the potential. Root mean square error (RMSE) in prediction of formation energy using MTP potential compared to one calculated using density functional theory (DFT) for training dataset is 2.1 meV/atom and for testing dataset is 6.8 meV/atom. The frequency of absolute error in formation energy peaks at a maximum value of 3.8 meV/atom for system that was part of training dataset, while it peaks at 7.6 meV/atom for systems that are not part of the training dataset.  Furthermore, the distribution and variability of elastic constants across compositions are systematically evaluated, revealing trends consistent with DFT benchmarks. The developed potential was used to predict energy of new phases in Ti-N system. We show that structures with N/Ti ratios ranging from 0 to 1 can be thermodynamically stable. A maximum deviation of 10 meV/atom from the convex hull plot of formation energy 0K was observed for a few system. 
\end{abstract}


\end{frontmatter}

\section{Introduction}
\noindent {Ti-N material system finds range of applications
.TiN and Ti/TiN multilayer coatings find application in aerospace industry~\cite{Wrblewski2021,SUN2020}, as biological implants~\cite{VanHove2015,Oliva2023} or in industrial applications like the drill bits~\cite{SaiKrishna2022}. In many electronic devices, Ti/TiN bilayer act as diffusion barrier layer that stops Cu from diffusing into SiO\textsubscript{2}~\cite{Kizil2004}. The atomistic simulations suggest the possibility of forming a range of Ti-N stoichiometry at the interface in Ti/TiN heterostructure~\cite{Gollapalli2022}.Several compounds like Ti$_2$N, Ti$_3$N$_2$,  Ti$_4$N$_3$ along with Ti \& TiN and solid solutions of N in Ti are synthesized experimentally~\cite{wriedt1986binary} and some of them like Ti$_6$N$_5$ are predicted via Density Functional Theory~\cite{Yu2015}

\noindent{To get a better understanding of Ti-N system and to perform reliable simulation across the range of chemistries there is need to develop an interatomic potential. The current state of the art MEAM potential~\cite{Kim2008} while does a decent job at computing the cohesive energies of Ti \& TiN but struggles to calculate the energies of other intermediate compounds and solid solutions. In this regard, Machine Learning approach has been a proven technique to build fast, accurate and robust interatomic potentials that captures complex atomic interactions~\cite{Wan2024,Ran2024,Mueller2020,Qiao2020}}
\\Moment Tensor Potential (MTP) is one of them which has evolved into a well established class of interatomic potentials. A comparative analysis with other interatomic potential indicates that the MTP achieves lower errors in energy, forces and stress predictions, highlighting its improved accuracy. Wang et al.~\cite{Wang2025} applied MTP for Ni-Al systems and demonstrated that it predicts better than traditional semi-empirical potentials such as EAM, ADP and MEAM in predicting energies, forces and stresses. For Ni-Al systems, MTP achieved low energy prediction errors, with RMSE of 2.24meV/atom for training set and 1.64 meV/atom for the validation set. It also achieved a force RMSE of $0.029~\text{eV}/\text{\AA}^2$ for the training set and $0.028~\text{eV}/\text{\AA}^2$ for the validation set. Also, a comparative analysis on the Molybdenum systems, suggested that MTP demonstrates a lowest mean error and narrowest error distribution at temperatures as high as 2600K. Even stress errors are reliably reproduced by MTP in comparison to other potentials~\cite{Gubaev2023}. These findings highlights the robustness and predictive accuracy of MTP across wide range of thermodynamic system. While kim et al.~\cite{Kim2008} has already developed a MEAM potential for Ti-N and Ti-C system by fitting potential parameters to experimental data on the enthalpy of formation, lattice parameters, elastic constant and thermal linear expansion, the application is limited to Ti, Ti$_2$N and TiN. The ability to comprehensively describe bulk behavior is however limited. As reported by Prince et al.~\cite{Gollapalli2022}, there are several compounds like Ti$_2$N, Ti$_3$N$_2$, Ti$_6$N$_5$, Ti$_4$N$_3$ along with Ti and TiN and solid solutions like $\mathrm{Ti{-}0.14}\mathrm{N}$ \& $\mathrm{Ti{-}0.2}\mathrm{N}$ which forms crucial part of the Ti-N system. Modeling of the bulk system thus provides a foundational advantage when scaling up the simulations to more complex, real world systems involving the diverse Ti-N compounds and solutions. MLIP based on M3GNet was trained on Ti, TiN, Ti$_2$N. We cannot expect it to work for other Ti-N system as a machine learning model’s ability to accurately predict a property depends heavily on the training data it receives. We made efforts to understand the structure across the system to make right selection of training dataset.
\section{Computational Methods}
\subsection{Moment Tensor Potential}
Moment Tensor Potential or MTP are class of interatomic potentials introduced by Shapeev et al.~\cite{Shapeev2015}. It provides an accurate framework in modeling interatomic interactions in wide range of materials systems that includes elemental solids, compounds and alloys~\cite{Qi2023}. The core principle of MTP lies in the description of local atomic environment through set of moment tensors $M_{\mu, \nu}$, defined as follows: 
\begin{equation}
M_{\mu, \nu}(n_i) = \sum_j f_{\mu}(|\mathbf{r}_{ij}|, z_i, z_j) \, \mathbf{r}_{ij} \otimes \cdots \otimes \mathbf{r}_{ij}
\end{equation}

where n$_i$ captures the local neighborhood of atom i, including the relative positions of the neighboring atoms. The vector r$_{ij}$ points from atom i to atom j and z$_i$, z$_j$ represents the atomic types. These radial information are encoded in the function f$_\mu$. The angular informations are however captured as a tensorial product of the relative position vectors, producing a rank- $\nu$ tensor.

In our case, there are primarily two different species that are part of our study viz.. Titanium(Ti) and Nitrogen(N). To develop an effective interatomic potential for the different configurations of the Ti-N system, there are two crucial parameters that can govern the efficiency of MTP development: the cut-off radius r$_c$ and the maximum tensor level lev$_{max}$. The maximum tensor level was set to 22. This choice was made ensuring a appropriate compromise between computational efficiency and model complexity, which is high enough to capture the atomic environment features without unnecessarily increasing the computational cost. A total of 421 basis functions and 2621 moment tensor descriptors were utilized. The minimum and the maximum radial cut-off distance were fixed at 2\text{\AA} and 7\text{\AA} respectively in order to define the region where the polynomials can be scaled.
\section{Results}
\subsection{Density Functional Theory \& Molecular Dynamics Simulations}
The first principle DFT calculations~\cite{HOHENBERG}~\cite{Kohn} was performed using Vienna Ab Initio Simulation Package (VASP)~\cite{Kresse1996a},~\cite{Kresse1996b} which employs a plane wave basis set. The DFT calculations employed the Perdew, Burke and Ernzhoff (PBE)~\cite{Perdew1996a}, generalized gradient approximation (GGA), exchange correlational functional and the projector augmented method \cite{Joubert1999} for Titanium (Ti) and Nitrogen (N) element respectively. For all calculation, a plane wave kinetic energy cut off of 520 eV was used for the plane wave expansion of wave functions for calculations of highly accurate forces. The brillioun zone sampling was done using Monkhorst Pack Scheme~\cite{Monkhorst1976} with mesh resolution less than \( (2\pi \cdot 0.03\, \text{\AA}^{-1}) \).Structure relaxations were performed on Ti-N based compounds ie... Ti, Ti$_2$N, Ti$_3$N$_2$, Ti$_4$N$_3$, Ti$_6$N$_5$ and TiN using the conjugate gradient algorithm and the total energies converged to within \( 10^{-9} \) eV/atom.
\\In order to further evaluate the thermal stability and dynamic response of the different Ti-N structures, AIMD was performed subsequently. The simulations were performed using VASP with PBE-GGA exchange correlational functional[5] and PAW method. To ensure a more representative descriptions of the atomic interactions and surrounding environments, AIMD simulations were performed using enlarged supercell. For simulating the structures using realistic thermodynamic conditions, the NpT ensemble was chosen to maintain constant pressure while allowing the cell volume fluctuations. The NpT ensemble was employed using a Langevin thermostat(MDALGO=3, ISIF=3)~\cite{https://www.vasp.at/wiki/index.php/LANGEVIN_GAMMA}. The atomic and lattice friction coefficients set to $3ps^{-1}$. Equations of motion were integrated via the Verlet algorithm with  1fs time step. For Parinello-Rahman dynamics (MDALGO=3), PMASS was set to the total atomic mass of the simulation calculated as sum of indiviual atomic mass weighted by their stoichiometry in POSCAR. The Brillioun zone was sampled with Monkhorst-Pack K-point Mesh[15], with grid dimensions adjusted according to the supercell size. 

\subsection{Training Dataset Generation \& Dataset Selection}

\noindent{A comprehensive training dataset should represent the problem domain and encompass all relevant structures and possible composition to learn the complete structural and chemical complexity of the material system~\cite{Wang2024}. Various stable structure of Ti-N system,their space group and Wyckoff positions are listed in Table~\ref{tab:TiN_structures}.Structures like Ti, Ti$_2$N, Ti$_4$N$_3$, Ti$_3$N$_2$ and TiN have been  experimentally reported ~\cite{Stoehr2011}~\cite{Liu2012}~\cite{Yu2015}, whereas structure such as Ti$_6$N$_5$ has been predicted using DFT ~\cite{Yang2017} . While it may initially appear that various compounds belong to different space group, it should be noted that while TiN has cubic rock salt structure, the reported phases Ti$_6$N$_5$, Ti$_4$N$_3$ and Ti$_3$N$_2$ are versions of this structure with ordered N-vacancies. In Ti$_3$N$_2$,one third of Nitrogen sites are vacant, in Ti$_4$N$_3$ - one quarter and in Ti$_6$N$_5$ one sixth  This structural similarity allows a systematic selection of training structures~\cite{Yu2015}. Hence, Ti$_4$N$_3$, Ti$_3$N$_2$, Ti$_2$N, TiN, and Ti were included in the training dataset, while Ti$_6$N$_5$ was intentionally left out to test the robustness of the developed interatomic potential. Similarly, in the case of solid solutions where nitrogen atoms occupy octahedral voids in HCP-Ti to form compositions like Ti–0.14N, Ti–0.2N, and Ti–0.25N ( ranging from 17 to 23 at\%  depending on the temperature) ~\cite{Varalakshmi2025} , Ti–0.14N was used for training, while Ti–0.2N was excluded to assess the generalization of the model to solid solutions configurations. Such strategy serves as an effective way to generalize the model across different structures and make reliable predictions for unknown compositions.
Bulk structures at 0k were included as part of the training dataset. 
Also in order to ensure the accuracy of elastic constants predictions, strained structures were used.
\begin{table}[!htbp]
\centering
\caption{\textbf{The space group, lattice parameters, and Wyckoff positions for various Ti--N phases. In Ti$_x$N, $x$ is the percentage of N in Ti as an interstitial solid solution, where $0 < x \leq 0.25$.}}
\vspace{10pt}
\label{tab:TiN_structures}
\scriptsize
{\scriptsize
\renewcommand{\arraystretch}{1.4}
\begin{tabular}{llcccccc|lllll}
\hline
\textbf{Structure} & \textbf{System} & $a$ & $b$ & $c$ & $\alpha$ & $\beta$ & $\gamma$ & \textbf{Atom} & \textbf{Site} & $x$ & $y$ & $z$ \\
 &  & (\AA) & (\AA) & (\AA) & ($^\circ$) & ($^\circ$) & ($^\circ$) & & & & & \\
\hline
TiN & cubic (Fm$\bar{3}$m) & 4.252 & 4.252 & 4.252 & 90 & 90 & 90 & Ti & 4a & 0 & 0 & 0 \\
 & & & & & & & & N & 4b & 0.5 & 0.5 & 0.5 \\
Ti$_6$N$_5$ & monoclinic (C2/m) & 5.214 & 9.015 & 8.517 & 90 & 144.8 & 90 & Ti & 8j & 0.487 & 0.824 & 0.746 \\
 & & & & & & & & Ti & 4i & 0.001 & 1 & 0.738 \\
 & & & & & & & & N & 4h & 1 & 0.335 & 0.5 \\
 & & & & & & & & N & 4g & 0.5 & 0.166 & 0 \\
 & & & & & & & & N & 2c & 1 & 0 & 0.5 \\
Ti$_4$N$_3$ & monoclinic (C2/m) & 9.934 & 3.018 & 10.302 & 90 & 150.3 & 90 & Ti & 4i & 0.266 & 0.5 & 0.633 \\
 & & & & & & & & Ti & 4i & 0.258 & 0 & 0.885 \\
 & & & & & & & & Ti & 4i & 0.001 & 0.5 & 0.749 \\
 & & & & & & & & N & 2a & 0 & 1 & 1 \\
Ti$_3$N$_2$ & orthorhombic (Immm) & 4.156 & 3.035 & 9.16 & 90 & 90 & 90 & Ti & 2d & 0.5 & 0 & 0.5 \\
 & & & & & & & & Ti & 4j & 0 & 0.5 & 0.316 \\
 & & & & & & & & N & 4f & 0 & 0.5 & 0.162 \\
Ti$_2$N & tetragonal (P4$_2$/mnm) & 4.956 & 4.956 & 3.036 & 90 & 90 & 90 & Ti & 4f & 0.203 & 0.797 & 0.5 \\
 & & & & & & & & N & 4f & 0 & 0 & 0 \\
Ti--xN & hexagonal (P6$_3$/mmc) & 2.935 & 2.935 & 4.647 & 90 & 90 & 120 & Ti & 2c & 0.333 & 0.666 & 0.25 \\
 & & & & & & & & N(x) & 2a & 0 & 0 & 0 \\
Ti & hexagonal (P6$_3$/mmc) & 2.935 & 2.935 & 4.647 & 90 & 90 & 120 & Ti & 2c & 0.333 & 0.666 & 0.25 \\
\hline
\end{tabular}
}
\end{table}
In addition to that, AIMD data upto 300K ensured an effective way to capture the interactions at room temperature. 
\\Furthermore, beyond the structural diversity, an additional context in the selection of the training dataset is inspired by the work of Shuyin et al ~\cite{Yu2015}. The study reports that Ti$_6$N$_5$ and Ti$_4$N$_3$ adopt C2/m structure. This structural similarity leads to systematic approach in selecting structures for training. Thus from the reported phases, Ti$_4$N$_3$, Ti$_3$N$_2$ were selected in the training dataset along with TiN, Ti$_2$N and Ti respectively. Ti$_6$N$_5$ is deliberately excluded from the training dataset so that the model can be tested on it to check how it performs for a compound it hasn't been trained before. Such strategy serves as an effective way to generalize the model across different structures and make reliable predictions for unknown compositions. As an extension for Ti-N based solutions,$\mathrm{Ti{-}0.14}\mathrm{N}$ was included in the training dataset and $\mathrm{Ti{-}0.2}\mathrm{N}$ was in testing dataset to test the model's generalizability for solid solutions also. The various structures within the training as well as the testing dataset were subjected to different strain modes depending on their underlying crystal symmetry. For instance, TiN is the cubic crystal system, only three strain modes are required to determine the elastic constants using \textbf{(See Supplementary Information ,Table 1, Section 2)}. Similarly the strain modes for other crystal system are tabulated in \textbf{(See Supplementary Information)}. Application of systematic strains along different crystallographic directions in all these compounds and solutions ensured a wide range of deformation modes. The atomic environment with necessary structural diversity was also captured in the process. The analysis was initiated with the application of systematic strains to the equilibrium lattice structure of these compounds and solutions. In the given input, the number of strain cases is 7 and the corresponding strain values are: -0.015, -0.010, -0.005, 0.000, +0.005, +0.010 and +0.015. As reported in Table~\ref{tab:structure_counts}, all the strained structures generated were completely included in the dataset. 
\begin{table}[htbp]
\centering
\begin{tabular}{|c|c|c|}
\hline
\textbf{structure type} & \textbf{structure name} & \textbf{number of structures} \\
\hline
\multirow{6}{*}{Strained Structure} 
  & Ti & 124 \\
  & Ti$_2$N & 212 \\
  & TiN & 39 \\
  & Ti$_3$N$_2$ & 442 \\
  & Ti$_4$N$_3$ & 378 \\
  & $\mathrm{Ti{-}0.14}\mathrm{N}$ & 724 \\
\hline
\end{tabular}
\end{table}

\begin{table}[htbp]
\centering
\begin{tabular}{|c|c|c|c|}
\hline
\textbf{structure type} & \textbf{structure name} & \textbf{Initial count} & \textbf{Reduced count} \\
\hline
\multirow{5}{*}{Finite Temperature Structures} 
  & Ti & 2297 & 94 \\
  & Ti$_2$N & 2604 & 132 \\
  & TiN & 2000 & 45 \\
  & Ti$_3$N$_2$ & 1653 & 50 \\
  & Ti$_4$N$_3$ & 1944 & 158 \\
\hline
\end{tabular}
\caption{Structure counts for strained and high-temperature datasets}
\label{tab:structure_counts}
\end{table}

 Finite temperature calculations for Ti-N system involved simulations spanning from 2 to 13 ps to ensure configurational sampling. The temperature was applied from 10K to 400K to replicate the various thermodynamic states to reflect the operational condition observed in real-life scenarios. These structures were strategically selected to extract representative subset of the training dataset. One of the effective way of doing that is using Maxvol algorithm. Despite the comprehensive nature of this dataset, using it in its entirety for the model training is computationally expensive and time-consuming. To address this, a data selection algorithm called as Maxvol algorithm is used to extract a representative subset of the training data. Maxvol Algorithm is a known technique for identification of submatrix with maximum volume from a larger matrix ~\cite{Mikhalev2018}. In this technique only those data points are selected that captures the diversity of various atomic structures. The same was utilized for various training structures of Ti-N system.  The initial and the reduced counts of the structures of Ti-N based compounds is summarized in Table 2. The reduced count is considered in the training dataset. Overall 2398 structures were provided as input for the  potential development.

\subsection{Hyperparameter Tuning}
\noindent{The key hyperparameters considered were related to the weights assigned to energy, force and stress contributions to the loss functions. The weights were systematically tuned and a specific nomenclature was assigned to each of them as MTP$_{\mathit{e\_w{-}f\_w{-}s\_w}}$,where e\_w, f\_w and s\_w are the variables representing energy weights, force weights and stress weights respectively. Three different MTP potentials were defined in accordance to this:}
\\(i) {MTP$_{1{-}0{-}0}$}: It is selected since most of the calculations strictly depends on energy, priortizing energy the other weights ie...force and stress weights were kept 0
\\(ii) {MTP$_{100{-}10{-}1}$}: The choice was guided by the work of  Novoselov et al.[18], which reported lowest RMSE error for this.
\\(iii) {MTP$_{0.75{-}0.25{-}0.25}$}: Here, energy was assigned a comparatively higher value while the weights for force and stress were set equal.  The idea behind this choice was to maintain a balanced contribution from force and stress while prioritizing energy to capture fundamental stability of the structure.
\subsection{Validation of Interatomic Potential}
\noindent{The accuracy and transferability of interatomic potential trained with various hyperparameter for Ti-N system were evaluated with three statistical methods:}
\\(i) RMSE as a primary error metric to assess the potential performance by evaluation of rmse error of various interatomic potential against DFT reference data for energy, force and stress components.
\\(ii)Kernel Density Estimation (KDE) is used to assess the distribution profile of errors in the formation energy calculated by DFT and by the interatomic potential.
\\(ii)Box plot analysis along with swarm plot was used to visualize the spread of data points corresponding to each elastic constant across different Ti-N phases for different potentials respectively.

\subsubsection{Benchmarking Of Interatomic Potentials with RMSE metrics }
\begin{figure}[H]
  \centering
  \includegraphics[width=1.0\textwidth]{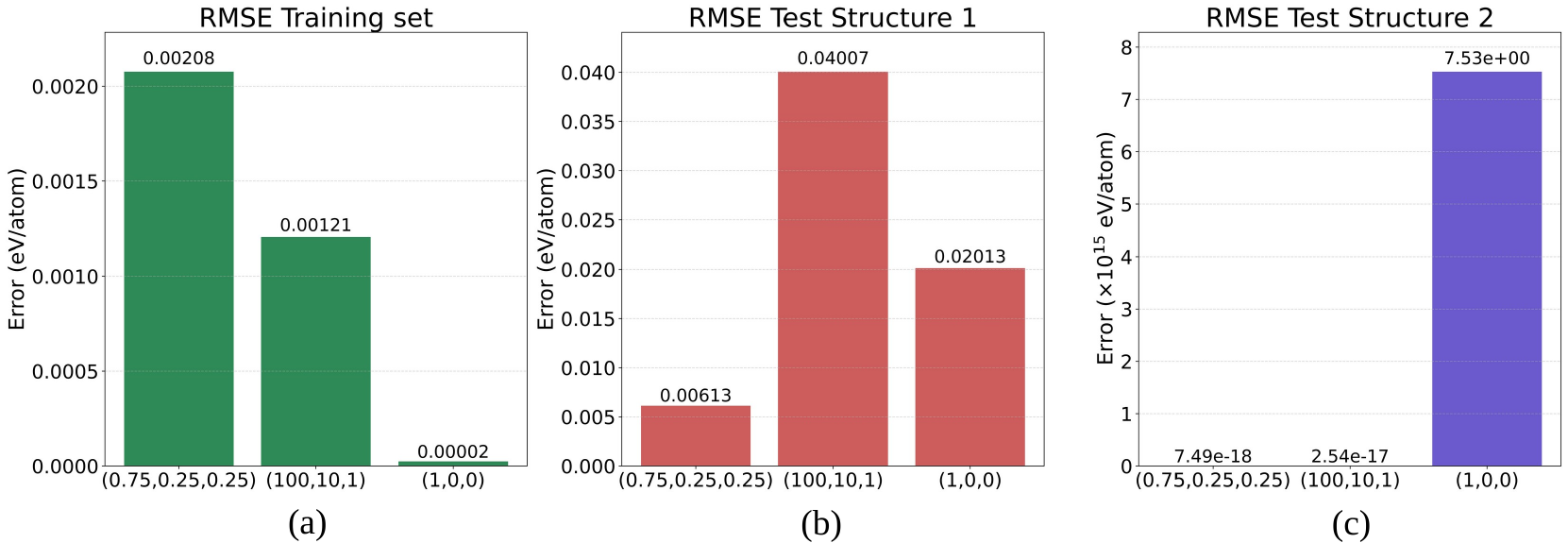}
  \caption{\textbf{Comparison of Root Mean Square Error (RMSE) values to assess the performance of interatomic potential across (a) Training dataset (b) Test Structure 1 (c) Test Structure 2}}
  \label{fig:error_metrics_potential}
\end{figure}
\noindent{To understand the efficacy of various interatomic potential, we calculate the RMSE in formation energy per atom with respect to DFT values for training dataset and also for structures which are not part of the training dataset ie.. the test dataset. The RMSE of training dataset and test dataset are plotted in Figure\ref{fig:error_metrics_potential}. From Figure\ref{fig:error_metrics_potential}, it is evident that {MTP$_{1{-}0{-}0}$} shows lowest RMSE for formation energy per atom in comparison to {MTP$_{100{-}10{-}1}$} and {MTP$_{0.75{-}0.25{-}0.25}$} for training dataset. However,for structures which are not part of the training dataset, the potential {MTP$_{100{-}10{-}1}$} and {MTP$_{0.75{-}0.25{-}0.25}$} shows comparatively lower RMSE values in comparison to {MTP$_{1{-}0{-}0}$}. These preliminary insights suggests that certain force and stress weights are equivalently important, even though the computational analysis primarily depends on energy predictions. This is crucial for the accuracy and transferability of the interatomic potential across various Ti-N structures. Thus, a comprehensive statistical analysis is essential to evaluate the effect of force and stress weight on the accuracy of energy prediction.}

\subsubsection{Analysis of Error in Energy Prediction}
To extensively evaluate the performance of different interatomic potential, the data points representing the absolute error between DFT formation energy values and the formation energy values predicted by each of the interatomic potential were compiled. This process was repeated for different structures of Ti-N system. The compiled errors from each system is examined with a statistical framework which conveys the probability distribution of data. This statistical framework is known as Kernel Density Estimation (KDE).
\\Kernel Density Estimation (KDE) is a non-parametric method used to estimate the probability density distribution of continuous variable through finite set of data points. In this technique, each data point corresponding to absolute error x$_i$, contributes to the probability density at a given target point x ,where the density is to be evaluated. The contribution of each error point is at that point is given by the gaussian kernel formula, scaled by bandwidth as follows:
\vspace{10pt} 
\begin{equation}
K\left( \frac{x - x_i}{h} \right) = \frac{1}{\sqrt{2\pi}} \exp\left( -\frac{(x - x_i)^2}{2h^2} \right)
\end{equation}
where x\textsubscript{i} denotes the error point, x denotes the points where the density is to be estimated and h is the bandwidth. The values of x is arbitrarily selected based on smoothening the final curve. Typically 100-200 values of x are selected within a specific range which is slightly lower than the minimum value of the error points and slightly higher than the maximum value of the error points.The bandwidth (h) that controls the width of each kernel is determined by the scott rule. The Scott's Rule for bandwidth selection in Kernel Density Estimation is given by the formula:
\vspace{10pt} 
\begin{equation}
h = \sigma \cdot n^{-1/5}
\end{equation}
The final KDE function is computed as the average of all kernel functions centered at each data point x\textsubscript{i}, scaled by bandwidth h. The formula is given by:
\vspace{10pt} 
\begin{equation}
\hat{f}(x) = \frac{1}{n h} \sum_{i=1}^{n} K\left( \frac{x - x_i}{h} \right)
\end{equation}
The factor $\frac{1}{nh}$ is multiplied for normalization, to ensure that the total area under all the kernel functions sums up to 1.
\\An effective utilization of KDE concept were used to evaluate different interatomic potentials across multiple structures. For the training part, structures Ti, Ti$_2$N, Ti$_3$N$_2$, Ti$_4$N$_3$, TiN \& $\mathrm{Ti{-}0.14}\mathrm{N}$ were employed, while for the validation part involved testing on unseen structures Ti$_6$N$_5$ \& $\mathrm{Ti{-}0.2}\mathrm{N}$. The performance of each interatomic potential under training and validation dataset is detailed as follows:
Error in prediction of energy compared to DFT of various structures used for training potential was calculated. Error for various structures and the kernel density estimation (KDE) plot is constructed by utilizing the error values between DFT reference values and the prediction from different interatomic points viz.. {MTP$_{1{-}0{-}0}$}, {MTP$_{100{-}10{-}1}$} and {MTP$_{0.75{-}0.25{-}0.25}$}. The KDE concept is built on kernel function, which is typically a gaussian at each indiviual error point x\textsubscript{i}. The Kernel is a non-parametric way of estimation of probability density function of a dataset . Probability density estimation is important because it helps understand the underlying distribution of a continous data without assuming that the curve is normal, uniform etc... This matters because it tells where the data is concentrated and where the data is sparse, whether the distribution is skewed or multimodal (more than one peak). It thus helps in understanding the true distribution. 
then KDE plots were generated for each of those structures with probability density versus absolute error data points as shown in Figure~\ref{fig:kde plot_train} \&~\ref{fig:kde plot_test}. The probability density corresponded to relative likelihood of the specific absolute error value within the dataset. Rather than indicating the probability density of individual values, it reflects how densely the absolute error values are concentrated at that point. A higher peak in the KDE curve signifies that the majority of the absolute errors for the structures are concentrated around that specific value. In combination with the peak, the spread of data contributed a major role. A diverse spread indicated greater variability/ inconsistency in prediction accuracy which is captured by FWHM (Full Width Half Maxima) of the peak.
\renewcommand{\arraystretch}{1.1} 
\begin{table}[h!]
\centering
\begin{tabular}{|>{\centering\arraybackslash}p{2.4cm}|>{\centering\arraybackslash}p{2.2cm}>{\centering\arraybackslash}p{2.4cm}|
>{\centering\arraybackslash}p{2.2cm}>{\centering\arraybackslash}p{2.4cm}|
>{\centering\arraybackslash}p{2.2cm}>{\centering\arraybackslash}p{2.4cm}|}
\hline
\textbf{Structures} & \multicolumn{2}{c|}{\textbf{MTP$_{1{-}0{-}0}$}} & \multicolumn{2}{c|}{\textbf{MTP$_{100{-}10{-}1}$}} & \multicolumn{2}{c|}{\textbf{MTP$_{0.75{-}0.25{-}0.25}$}} \\
\cline{2-7}
                    & \textbf{FWHM} & \textbf{PEAK\_POS} & \textbf{FWHM} & \textbf{PEAK\_POS} & \textbf{FWHM} & \textbf{PEAK\_POS} \\
\hline
Ti         & 0.1603 & 0.5861  & 1.0313     & 3.8263   & 1.1289     &  1.3317    \\
Ti$_2$N    & 0.209    & 0.2725    & 0.6899     & 0.3701     & 0.3414     & 0.5303     \\
TiN        & 0.0766     & 0.3561    & 0.4041     & 1.4989     & 0.4181     & 3.7915     \\
Ti$_3$N$_2$ & 0.1324   & 0.3979   & 0.2718     & 1.3108     & 0.2229     & 2.1957     \\
Ti$_4$N$_3$ & 0.2857   & 0.8230    & 0.2578     & 0.1819     & 0.3484     & 1.1366     \\
Ti-0.14N   & 2.091E+018   & 4.063E+018    & 0.1394   & 0.5094     & 0.1185     & 0.5303     \\
\hline
\end{tabular}
\caption{\textbf{Comparison of FWHM and PEAK\_POS for various trained structures under multiple MTP potentials}}
\label{tab:mtp_comparison_train}
\end{table}

\par

\begin{table}[h!]
\centering
\begin{tabular}{|>{\centering\arraybackslash}p{2.4cm}|>{\centering\arraybackslash}p{2.2cm}>{\centering\arraybackslash}p{2.4cm}|
>{\centering\arraybackslash}p{2.2cm}>{\centering\arraybackslash}p{2.4cm}|
>{\centering\arraybackslash}p{2.2cm}>{\centering\arraybackslash}p{2.4cm}|}
\hline
\textbf{Structures} & \multicolumn{2}{c|}{\textbf{MTP$_{1{-}0{-}0}$}} & \multicolumn{2}{c|}{\textbf{MTP$_{100{-}10{-}1}$}} & \multicolumn{2}{c|}{\textbf{MTP$_{0.75{-}0.25{-}0.25}$}} \\
\cline{2-7}
                    & \textbf{FWHM} & \textbf{PEAK\_POS} & \textbf{FWHM} & \textbf{PEAK\_POS} & \textbf{FWHM} & \textbf{PEAK\_POS} \\
\hline
Ti$_6$N$_5$ & 3.6492   & 20.0278   &  1.2301   &  40.1603    &  1.2711   &  5.9228   \\
Ti-0.2N   &  1.8721E+018 & 8.2722E+018   & 0.1640  & 25.3992    & 0.7381    & 7.5629    \\
\hline
\end{tabular}
\caption{\textbf{Comparison of FWHM and PEAK\_POS for various test structures under multiple MTP potentials}}
\label{tab:mtp_comparison_test}
\end{table}

\par
\noindent(i) \textbf{MTP$_{1{-}0{-}0}$ interatomic potential:} The interatomic potential was developed by prioritizing the energy as the reference weight. To reflect its importance the value of the energy parameter was set to 1 while remaining weight parameters ie...force and stress were set to zero. From Table~\ref{tab:mtp_comparison_train}, the width of the curve corresponding to MTP$_{1{-}0{-}0}$ interatomic potential shows peak position between 0.3 to 0.8 meV/atom and FWHM value between 0.1 to 0.3 meV/atom, suggesting high prediction accuracy and consistency. This works well for all the Ti-N compounds that has been part of training data. However, when extended to the testing set of structures, the error distribution became noticeably flatter for Ti$_6$N$_5$ (refer figure~\ref{fig:kde plot_test}). The FWHM comes out to be 3.6492 meV/atom and the peak position is at 20.0278 meV/atom. Also, in case of solid solution $\mathrm{Ti{-}0.14}\mathrm{N}$ (which is also the part of training data), the peak position as well FWHM is in orders of $10^{18}$. The same also holds true for testing data on $\mathrm{Ti{-}0.2}\mathrm{N}$ solid solution. Given the range and inconsistency of the observed errors in MTP$_{1{-}0{-}0}$ interatomic potential, this potential is not adequate to serve as a reliable model across broad range of Ti-N system
\begin{figure}[H]
  \centering
  \includegraphics[width=0.9\textwidth]{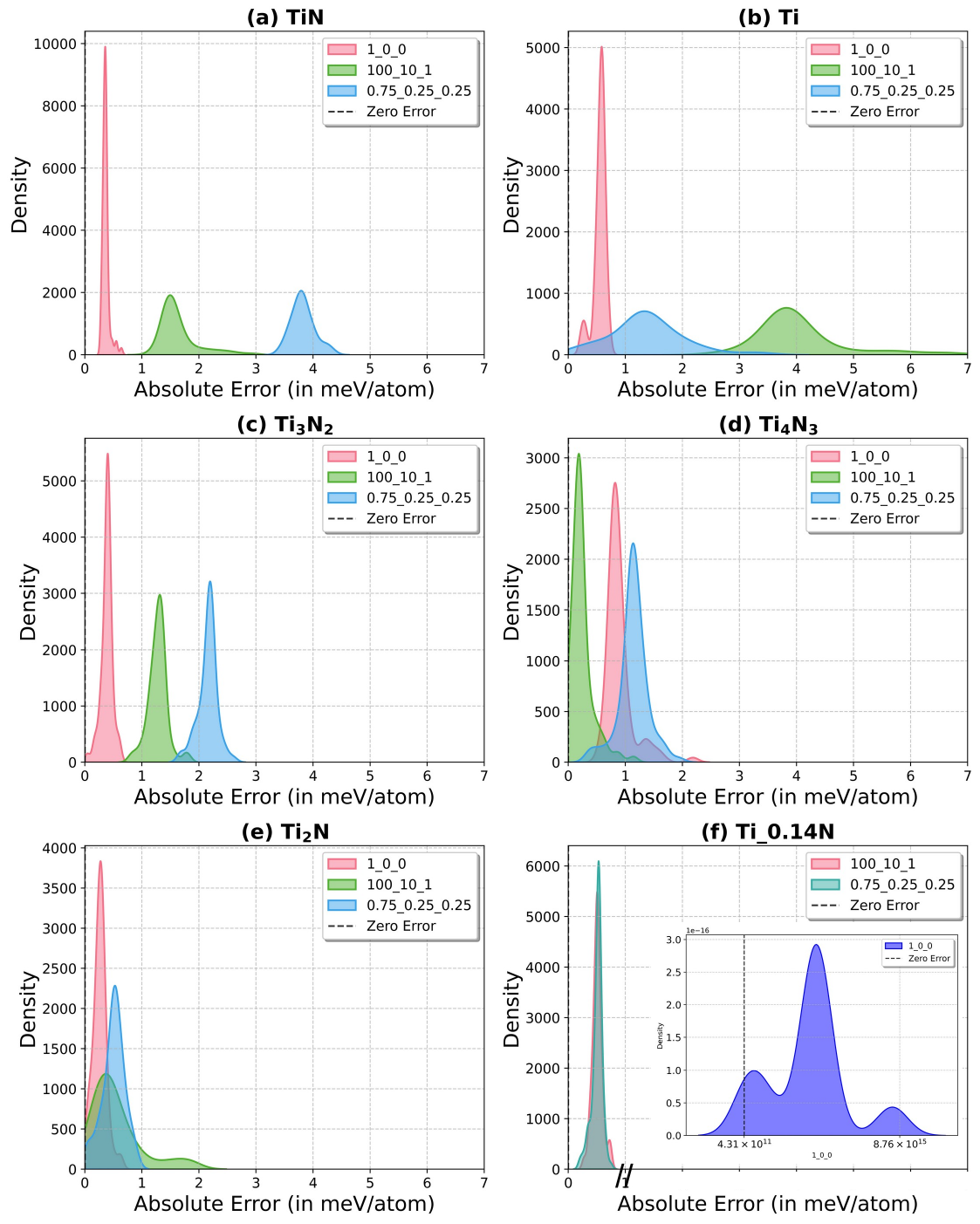}
  \caption{\textbf{Kernel Density Estimation (KDE) plots of the energy distributions for different structures of the training data in Ti-N system (a) TiN (b)Ti (c)Ti$_{3}$N$_{2}$ (d)Ti$_{4}$N$_{3}$ (e)Ti$_{2}$N (f)$\bm{\mathrm{Ti}\text{-}0.14\mathrm{N}}$. Each plot represent the absolute error distribution of different interatomic potentials w.r.t DFT calculated values}}
  \label{fig:kde plot_train}
\end{figure}

\begin{figure}[H]
  \centering
  \includegraphics[width=0.9\textwidth]{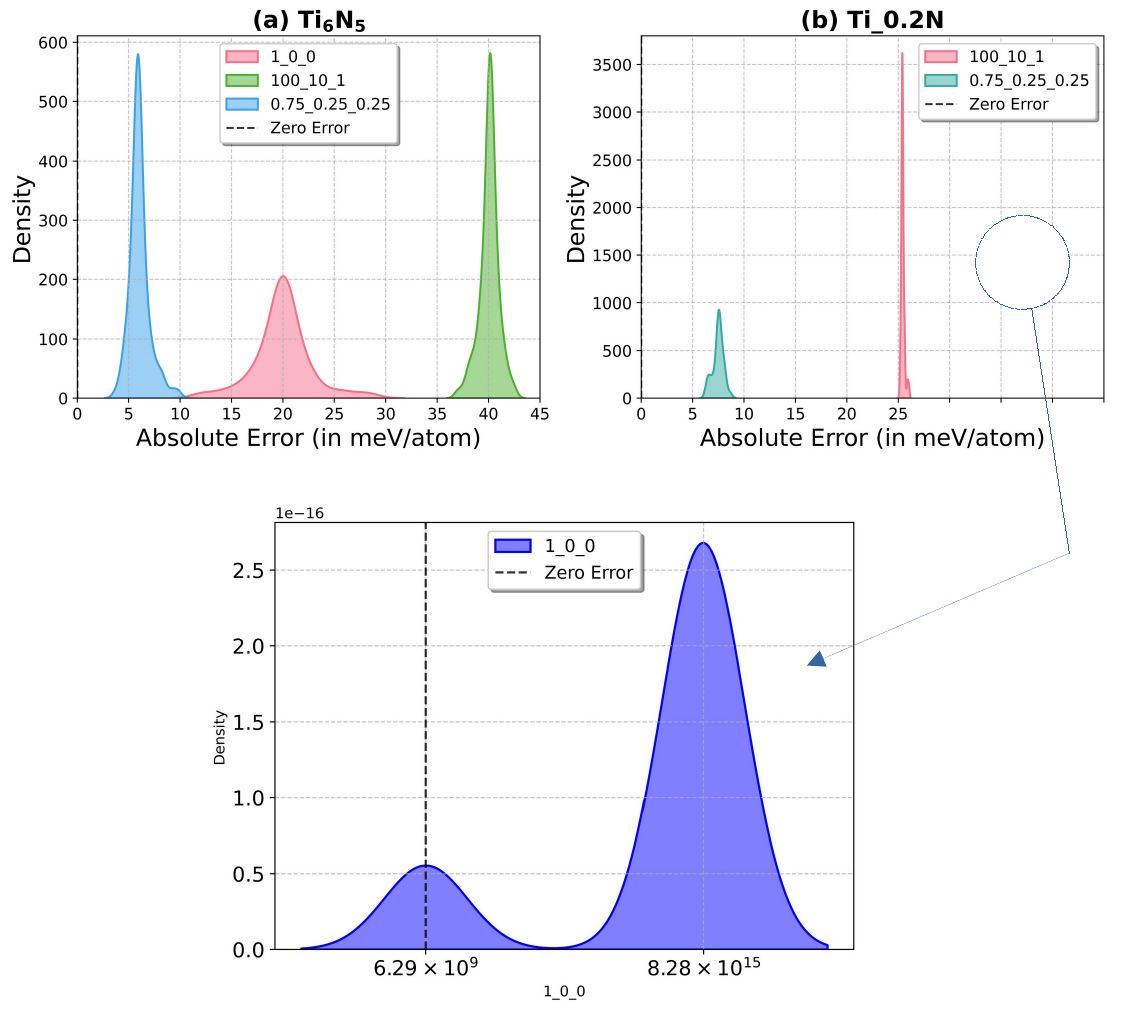}
  \caption{\textbf{Kernel Density Estimation (KDE) plots of the energy distributions for different structures of the test data in Ti-N system (a)Ti$_{6}$N$_{5}$ (b)$\bm{\mathrm{Ti}\text{-}0.2\mathrm{N}}$}}
  \label{fig:kde plot_test}
\end{figure}

(ii) \textbf{MTP$_{100{-}10{-}1}$ interatomic potential:} The selection of this interatomic potential was guided by prior results from the work of Novoselov et al.~\cite{Novoselov2019}. The interatomic potential reported low RMSE values, suggesting  high predictive accuracy. The energy weight was set to 100, while force and stress weight were set to 10 and 1 respectively. From Table~\ref{tab:mtp_comparison_train}, the FWHM of the curve corresponding to MTP$_{100{-}10{-}1}$ shows variation from 0.1 to 1 meV/atom and Peak position from 0.1 to 4 meV/atom for different range of Ti-N compounds as well as solution which were part of the training data. 
However, the error distribution is slightly higher for Ti. The deviation from the DFT data or the peak position is at 3.82 meV/atom as shown in Table~\ref{tab:mtp_comparison_train}. Building on this, we proceed to evaluate the potential on the validation dataset. 
From Figure~\ref{fig:kde plot_test}, it is observed that the error distribution in Ti$_6$N$_5$ is quite narrow and the peak is sharp. However, the deviation from the dft value or the peak position is at 40.16 meV/atom which is abysmally high. In addition, the solid solution $\mathrm{Ti{-}0.2}\mathrm{N}$, on which the potential is evaluated also which shows the peak position at 25.4meV/atom. 
While this potential provides a marginal improvement over the previous one, further assessment with yet another interatomic potential would provide a clear picture of the predictive performance.

\noindent(iii) \textbf{MTP$_{0.75{-}0.25{-}0.25}$ interatomic potential:} The analysis of preceding two interatomic potential necessitates the importance of incorporating both the stress and the force components in appropriate proportions to enhance the model's overall reliability. This approach ensures balanced contribution from force and stress while prioritizing energy to capture fundamental stability of the structure. The energy weight was set to 0.75, force weight to 0.25 and stress weight to 0.25 respectively. From Figure~\ref{fig:kde plot_test}, it is clear that the FWHM corresponding to MTP$_{0.75{-}0.25{-}0.25}$ shows variation from 0 to 4 meV/atom from across the Ti-N system, both for compounds as well as solid solution. Continuing the assessment, the potential is tested on the validation dataset to evaluate its generalization. It is observed that the error distribution in Ti$_6$N$_5$ is narrow, with sharp peak and is near to the DFT value (Refer Figure~\ref{fig:kde plot_test}). The peak position for Ti$_6$N$_5$ is at 5.9228 meV/atom which is much better than the previous two potentials. Similarly, for  $\mathrm{Ti{-}0.2}\mathrm{N}$ , peak position is at 7.6 meV/atom, which is low and aligns well with our objectives. In addition to this the FWHM value for both Ti$_6$N$_5$ and $\mathrm{Ti{-}0.2}\mathrm{N}$ are 1.2711 meV/atom and 0.7381 meV/atom respectively which implies the distribution is sharply peaked.

Therefore, {MTP$_{0.75{-}0.25{-}0.25}$} stands out as the most reliable among all the tested potentials in terms of accuracy and consistency with respect to prediction of formation energy with respect to diverse strained and molecular dynamics configurations.

\subsubsection{Analysis of error in calculation of Elastic Constants}
Using the energy strain method, elastic constants were calculated. Here, each strain index from the strain vector ($\epsilon_x$,$\epsilon_y$,$\epsilon_z$,$\epsilon_{xy}$,$\epsilon_{yz}$$\epsilon_{zx}$) represents the deformation modes, where the first 3 terms corresponds to normal strains along principal axis and last three terms corresponds to shear strains along xy, yz and zx planes. In correlation to the above strains, the elastic energy ($\Delta E$) corresponds to how the system will interact under energetically different strains. The elastic energy ($\Delta E (V, {\epsilon_i})$) of a solid under strain in harmonic approximation (as reported by Wang et al.) is given by:

\vspace{10pt} 
\begin{equation}
\Delta E (V, {\epsilon_i}) = E(V, {\epsilon_i}) - E(V_o, 0) = \frac{V_0}{2} \sum_{i,j=1}^{6} C_{ij} \varepsilon_i \varepsilon_j
\end{equation}

As an example, for a cubic crystal under biaxial normal strain condition [$\epsilon_x$,$\epsilon_y$,$\epsilon_z$,$\epsilon_{xy}$,$\epsilon_{yz}$$\epsilon_{zx}$] = [$\delta$,$\delta$,0,0,0,0] and the corresponding elastic energy-deformation comes out as:

\vspace{10pt} 
\begin{equation}
\Delta E (V, {\epsilon_i}) = \frac{V_0}{2} (C_{11} + 2C_{12}) \delta^2
\end{equation}
The strain vectors in different crystal systems (Refer Appendix 1 of \cite{Wang2021}) demonstrate the deformation modes applied to the crystal system. For each of them, the elastic strain energy per unit volume was calculated and the value of elastic constants predicted through DFT and potential approach.  
\\While the energy based evaluations identified {MTP$_{0.75{-}0.25{-}0.25}$} as a reliable potential, a more comprehensive validation requires assessing the performance in prediction of elastic constants. The elastic constants serves as a fundamental parameters for determination of key mechanical properties including bulk modulus, shear modulus and Young's modulus. These properties governs the material's response to external stresses and are critical for applications involving mechanical stability and structural performance. Thus, evaluation of percentage errors in elastic constant predictions across different potentials provides a better insight into its transferability and accuracy.
\\To address this, a combination of box plot and swarm plot was employed. The box plot utilizes the elastic constants (eg.. C11, C22, C33 etc..) to evaluate the statistical distribution of percentage errors across different interatomic potentials as shown in Figure~\ref{fig:swarm_plots}. The selected elastic constants were specifically chosen based on their contribution in calculation of the mechanical properties like bulk modulus, elastic modulus and shear modulus. For the analysis in the distribution of percentage errors in elastic constants, the key percentiles used are: the 25th (Q\textsubscript{1}), 50th (median Q\textsubscript{2}) and 75th (Q\textsubscript{3}) percentiles. These percentiles tells the percentage of values which falls below a certain point in the dataset. For example, the 25th percentile implies 25\%  of values are lower than that value. It helps in understanding the spread and ranking of data. These statistical markers alongwith the interquartile range IQR defined as: 
\vspace{10pt} 
\begin{equation}
IQR = Q\textsubscript{3} - Q\textsubscript{1}
\end{equation}
captures 50\% of the data. We utilize this concept to evaluate the elastic constants. For each elastic constant, the percentage error is first computed as:

\vspace{10pt} 
\begin{equation}
E_i = \left( \frac{C_i^{\text{ref}} - C_i^{\text{pred}}}{C_i^{\text{ref}}} \right) \times 100
\end{equation}

and thereby a box plot is generated using  Q\textsubscript{1},  Q\textsubscript{2} and  Q\textsubscript{3} with whiskers extending up to 1.5 x IQR. The process is repeated across relevant elastic constants for a given potential. Median of these errors are used thereby calculated. For all potentials viz.. {MTP$_{1{-}0{-}0}$}, {MTP$_{100{-}10{-}1}$} and {MTP$_{0.75{-}0.25{-}0.25}$}, the process is repeated. The median error and IQR are calculated for each is used as a representative measure to compare the accuracy of different potentials. The one with lowest median error and IQR value is considered as the best potential.

Yet another metric to assess the overall bias in the potential is mean error(ME). Despite its sensitivity to outliers, \textbf{ME} is still a good indicator for evaluation of model performance. The ME is defined (in our context) as:
\vspace{10pt} 
\begin{equation}
\mathrm{ME} = \frac{1}{n} \sum_{i=1}^{n} \left( \frac{C_i^{\text{pred}} - C_i^{\text{ref}}}{C_i^{\text{ref}}} \times 100 \right)
\end{equation}

A low mean error (ME) indicates the low presence of outliers. Since ME is sensitive to outlier a large value of outlier can disproportionately effect the value. Thus, a low ME indicates minimal average deviation from reference values and is also suggestive of low presence of outliers.

\noindent The box plot captured the median error, whisker plot and the interquartile range (IQR) for each interatomic potential across various elastic constants. Each box represents the statistical spread of errors. Median error provides a robust central measure less sensitive to outliers, while IQR indicates the spread of middle 50\% of the error distribution. The mean error gives the overall average but may be influenced by outliers. In contrast, the swarmplot displays the distribution of individual data points in each category, highlighting variations and outliers. The raw percentage errors for each of the potential ie... {MTP$_{1{-}0{-}0}$}, {MTP$_{100{-}10{-}1}$} and {MTP$_{0.75{-}0.25{-}0.25}$} were collected across different elastic constants viz..C11, C22, C33, C44, C55, C66, C12, C13 and C23 respectively. The pooled raw percentage error values were computed to a single cumulative median error for that potential. Along with cumulative median error value, the same pooling approach was utilized for IQR and mean as well. The results of the same are in table~\ref{tab:potentials_table}.

\begin{table}[h!]
\centering
\caption{Summary of Interatomic Potentials}
\label{tab:potentials_table}
\begin{tabular}{|c|c|c|c|c|}
\hline
\textbf{Potential} & \textbf{Median Error (\%)} & \textbf{IQR (\%)} & \textbf{Mean Error (\%)} & \textbf{Rank} \\
\hline
{MTP$_{0.75{-}0.25{-}0.25}$} & 4.83 & 6.12 & 7.41 & 1   \\
{MTP$_{100{-}10{-}1}$} & 5.97 & 11.85 & 9.41 & 2 \\
{MTP$_{1{-}0{-}0}$} & 7.68 & 10.23 & 9.95 & 3 \\
\hline
\end{tabular}
\end{table}

From the table~\ref{tab:potentials_table}, potential was assessed based on three key statistical descriptors commonly associated with boxplots: median error, mean error and interquartile range. Based on the computation on these metrics, overall rank was associated to identify which potential provides the most accurate and consistent predictions of elastic constants across various Ti-N structures. It is seen that the Median error, IQR and the mean error for {MTP$_{0.75{-}0.25{-}0.25}$} is the lowest amongst all in terms of Percentage Median Error, IQR value and Mean Error. From Figure~\ref{fig:swarm_plots}, the swarmplot also reveals that {MTP$_{0.75{-}0.25{-}0.25}$} exhibits tighter clustering of percentage errors across various structures, indicating more consistent predictions. In addition to it, the percentage errors are also comparatively low, further supporting the superior performance in modeling of elastic constants.
\\Thus based on the combined analysis of the boxplot and swarmplot for the elastic constant percentage errors, it can be concluded that {MTP$_{0.75{-}0.25{-}0.25}$} demonstrates superior performance compared to other potentials.

\begin{figure}[H]
  \centering
  \includegraphics[width=0.9\textwidth]{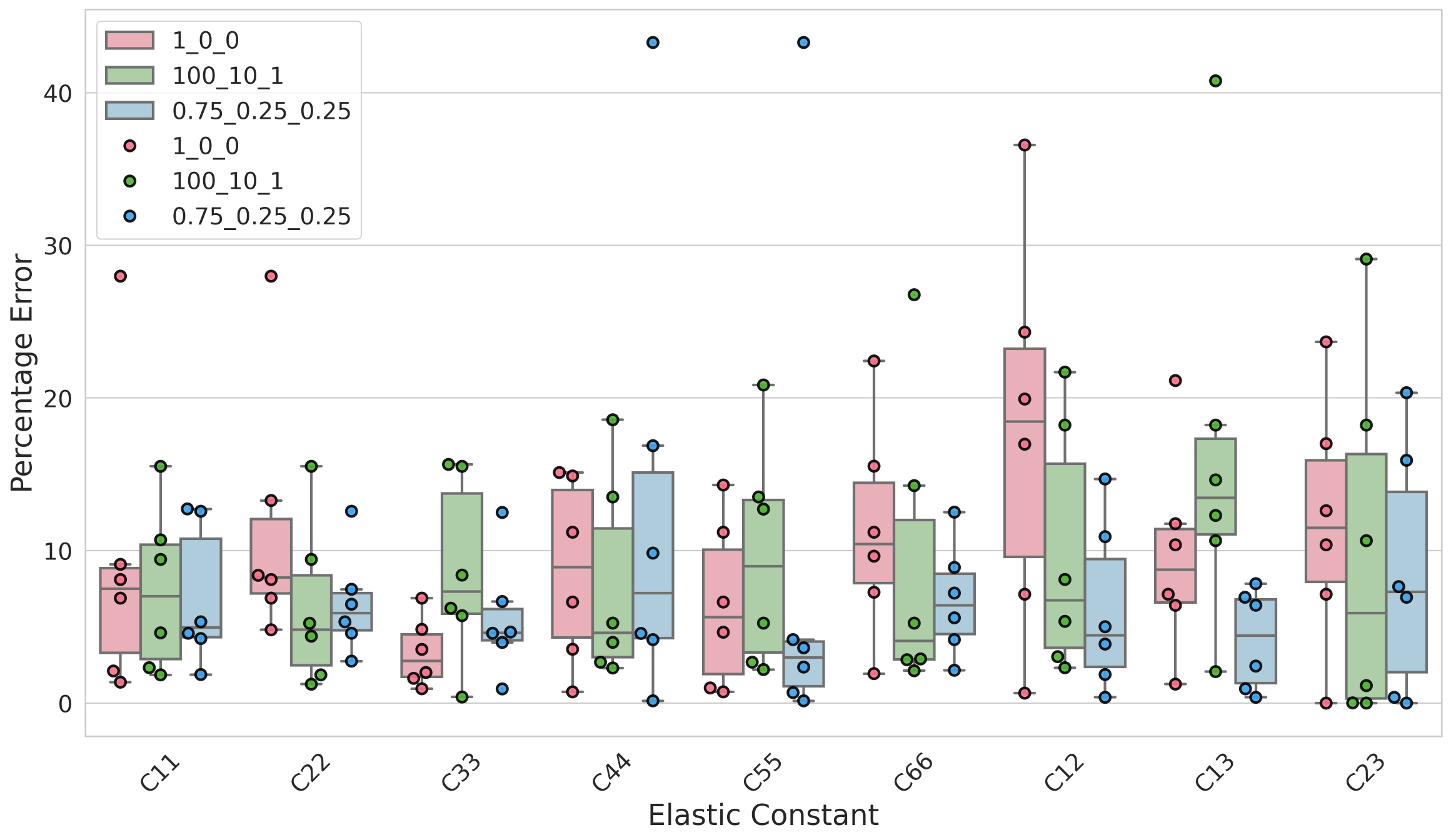}
  \caption{\textbf{Combined boxplot and swwarmplot analysis of percentage error distribution vs Elastic constants as predicted by various interatomic potentials. The visualization enables comparative analysis of predictive performance of potentials across different elastic constants}}
  \label{fig:swarm_plots}
\end{figure}

\subsection{MD simulation till room temperature}
Going forward we only consider MTP$_{0.75{-}0.25{-}0.25}$}. 
We evaluated the efficacy of the {MTP$_{0.75{-}0.25{-}0.25}$} to perform molecular dynamics of various Ti-N system using LAMMPS. Ti\textsubscript{2}N ,Ti\textsubscript{3}N\textsubscript{2}, Ti\textsubscript{4}N\textsubscript{3}, Ti\textsubscript{6}N\textsubscript{5} supercell containing 36, 72, 42, 44 and 64 atoms were heated from 10 K to 300 K. As illustrated in Figure~\ref{fig:temperature_time_plots}, the temperature vs time profiles has been plotted for different compounds which are part of the Ti-N system. The analysis of temperature-dependant fluctuations was used to assess the thermal behaviour and stability of the materials with NPT canonical ensemble. The integration time step was maintained at 1fs. Zero external pressure ensured that no external stress was applied. Temperature and pressure were controlled using Nose Hover Thermostat and Barostat with damping constants of 2ps and 10ps respectively. At low temperatures ie..at 10K, the system remains stable. However as the temperature increases the net fluctuation in various phases of Ti-N system are typically in range of (1\%-10\%) respectively, consistent with the expected equilibrium deviations in NPT simulations, as cited by Meharenna et al.~\cite{Meharenna2010}.
Analysis of the structure also indicates that there isn't any abrupt instability in the structures obtained at elevated temperature. 
\begin{figure}[H]
  \centering
  \includegraphics[width=1.0\textwidth]{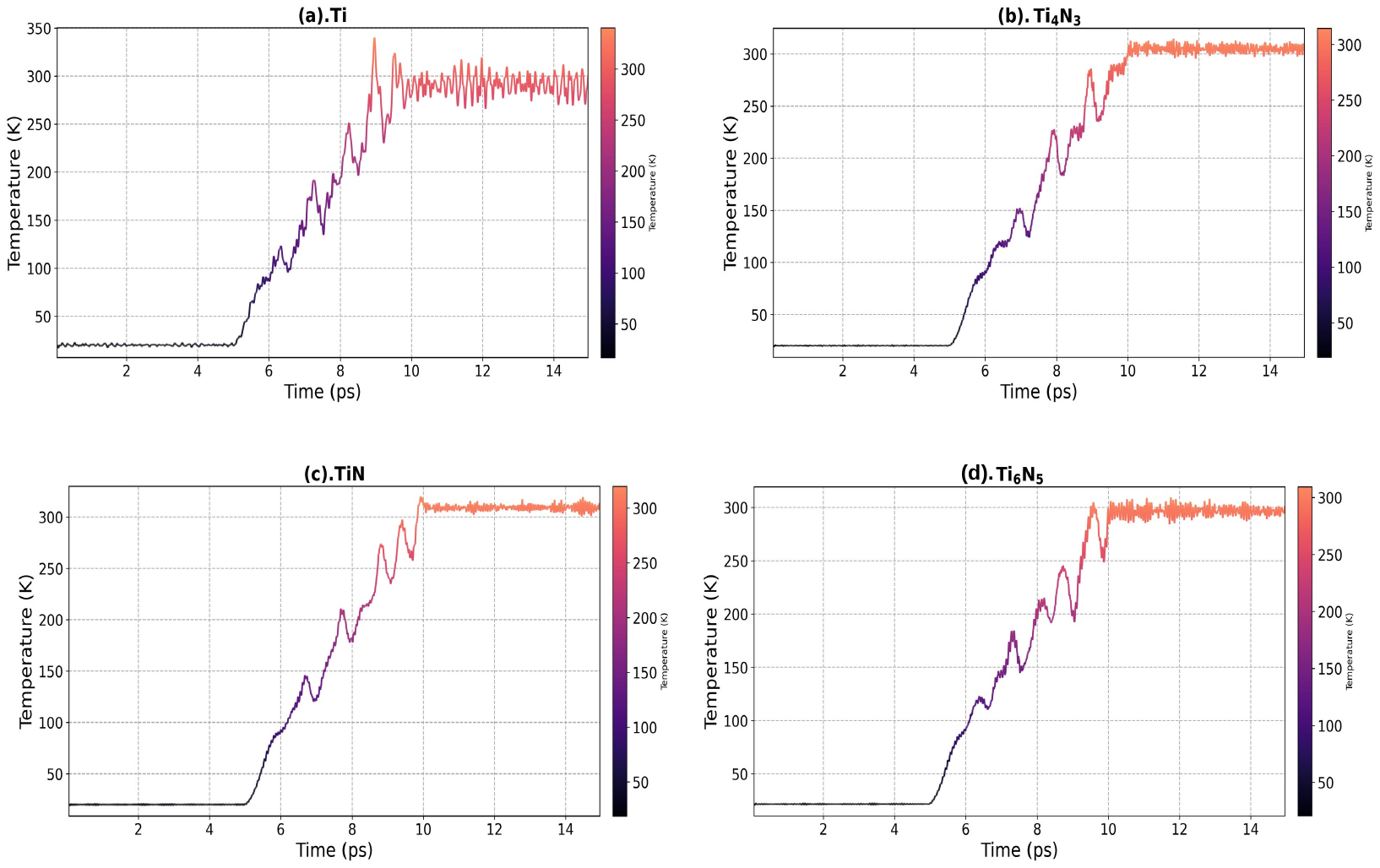}
  \caption{\textbf{Temperature vs time plot for different structures of Ti-N system(a) Ti (b) Ti\textsubscript{4}N\textsubscript{3} and (c) TiN are part of training dataset and (d) Ti\textsubscript{6}N\textsubscript{5} is part of test dataset. Each simulations were done to investigate the thermal response and structural stability as the temperature increases from 10K to 300K}}
  \label{fig:temperature_time_plots}
\end{figure}
\subsection{Comparison with other interatomic potentials}
To highlight the accuracy and consistency of various interatomic potentials viz... M3GNet ~\cite{Chen2022}, MEAM ~\cite{Kim2008} and {MTP$_{0.75{-}0.25{-}0.25}$} relative to DFT in order to capture the formation energy of different compounds and solutions of Ti-N system, we utilize the convex hull plot. The convex plot represents the formation energy of different Ti-N compounds as a function of composition. 
\begin{figure}[H]
  \centering
  \includegraphics[width=0.9\textwidth]{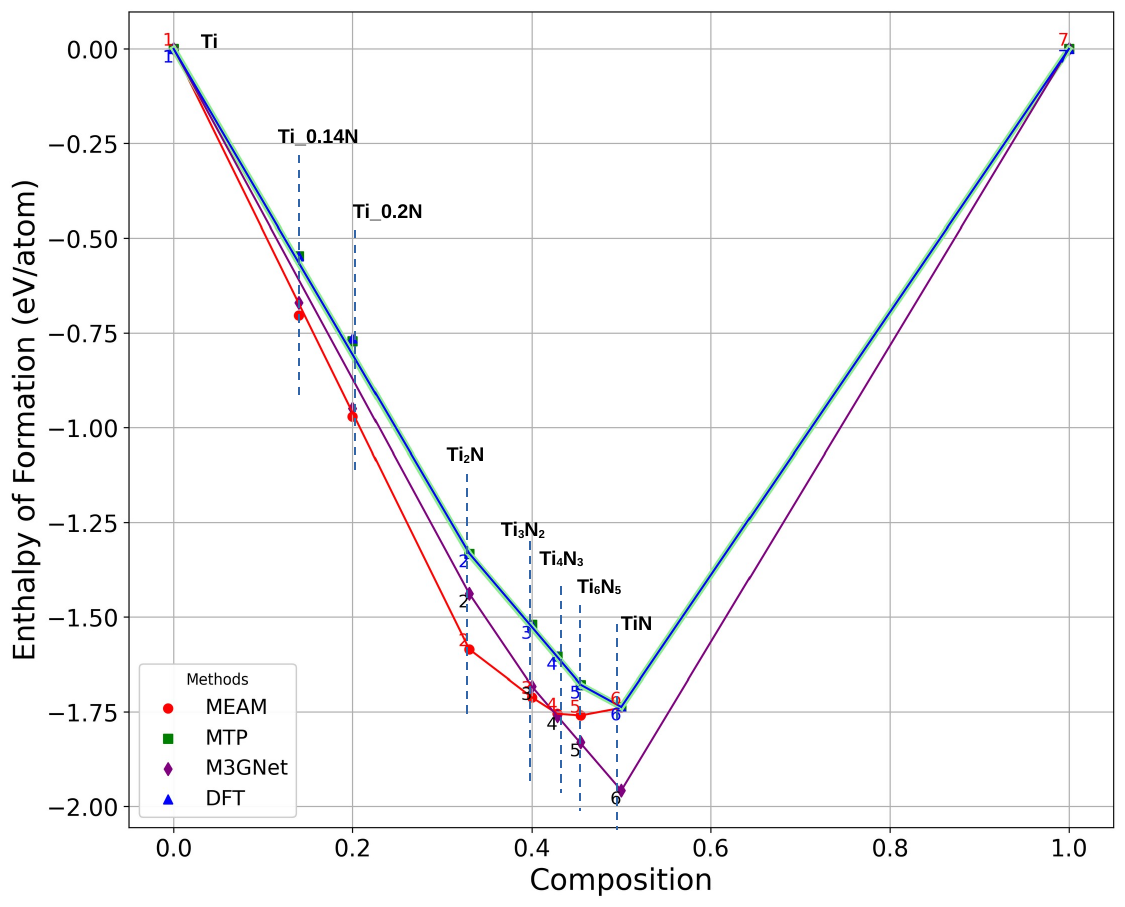}
  \caption{\textbf{Convex hull plot generated from different interatomic potentials for Ti-N system. The hull plots illustrates the phase stability as well deviation of hull from DFT predicted Convex hull}}
  \label{fig:convex_hull_plots_interatomic_potentials}
\end{figure}
The formation energies of these are calculated with the most stable reference states of Ti(hcp Ti) and N (N\textsubscript{2} gas). The equation that goes into calculation of the formation energy for compounds and solutions are as follows:
\begin{equation}
\Delta H^{f} (\mathrm{Ti}_{1-x}\mathrm{N}_x) = E(\mathrm{Ti}_{1-x}\mathrm{N}_x) - (1 - x)E(\mathrm{Ti}) - \frac{1}{2}x E(\mathrm{N}_2)
\end{equation}

\begin{equation}
\Delta H^{f} (\mathrm{Ti}_{1-x}\mathrm{N}) = E(\mathrm{Ti}_{1-x}\mathrm{N}) - E(\mathrm{Ti}) - \frac{1}{2}x E(\mathrm{N}_2)
\end{equation}

Where $E_{\mathrm{Ti}_{1-x}\mathrm{N}_x}$ and $E_{\mathrm{Ti}_{1-x}\mathrm{N}}$ are the DFT-calculated total energies of the pristine supercells with compounds and solid solutions respectively. $E(\mathrm{N}_2)$ is the DFT total energy of the N$_2$ molecule and $E(\mathrm{Ti})$ is the DFT total energy of Ti.

From Figure~\ref{fig:convex_hull_plots_interatomic_potentials}, the predicted hull plot from each potential is marked with distinct markers. The MEAM potential predicts values which qualitatively aligns with the DFT trend without overlapping with it. The deviation of the formation energy from DFT value for each compound and solution typically lies in the range of (80-250) meV/atom. Similarly, the deviation in the value of the formation energy calculated through M3GNet pretrained potential comes out to be in the range of (100-200) meV/atom. Given the typical variations in the energy predictions from the previous two potentials ie...MEAM potential and a pretrained M3GNet potential within 100s of meV/atom range, the focus of this work is on convex hull construction with different strained and MD structure as the part of the training data.
\\However, with the trained structures based  {MTP$_{0.75{-}0.25{-}0.25}$} potential, higher precision was obtained in reproducing the convex hull. It showed a near to perfect overlap with the DFT results as shown in Figure~\ref{fig:convex_hull_plots_interatomic_potentials}. While this level of agreement is expected as we trained the potential with these structures, it nonetheless demonstrates the capability of the potential to capture diverse Ti-N based energy landscape. The degree of accuracy in the prediction of the formation energy for structure typically lies in the range of (0 to 1.5) meV/atom.

\subsection{Prediction of New Ti-N structures}
We calculated formation energy of stable and metastable phases in Ti–N system, with the ratio of N/Ti varying from 0 to 1, using  {MTP$_{0.75{-}0.25{-}0.25}$} potential. This was in order used to construct 0 K phase diagram of Ti-N system. Intermediate structures were generated by inserting nitrogen atoms at octahedral voids in Ti, Ti\textsubscript{2}N ,Ti\textsubscript{3}N\textsubscript{2}, Ti\textsubscript{4}N\textsubscript{3}, Ti\textsubscript{6}N\textsubscript{5} and by removing nitrogen atoms from TiN, Ti\textsubscript{6}N\textsubscript{5}, Ti\textsubscript{4}N\textsubscript{3}, Ti\textsubscript{3}N\textsubscript{2}, and Ti\textsubscript{2}N respectively. For each structures, the formation energies obtained via both insertion and removal were calculated and compared to determine the energetically most stable structure.
\vspace{1\baselineskip}
\\(i)For structures with N/Ti ratios ranging from 0 to 0.5, two set of structure with same N/Ti ratio were modeled. The model scenarios includes: 
\noindent \\a) Insertion of N-atoms at octahedral voids in Ti-HCP supercell containing 96 Ti atoms
\noindent \\b) Removal of N-atoms from Ti2N supercell containing 96 Ti and 48 N atom. 
\\The first nitrogen atom was randomly placed in an octahedral void, and subsequent N atoms were iteratively added to the farthest available octahedral void to minimize local clustering. A maximum of 48 N atoms were inserted in the process. Each generated structure was relaxed using the {MTP$_{0.75{-}0.25{-}0.25}$} potential, and formation energies per atom were computed. Structure was created by removing N atoms from Ti\textsubscript{2}N supercell containting 96 Ti, 48 N atoms. One nitrogen atom was randomly removed from the supercell, to create suprcell contaning 96 T and 47 N atoms, atomic position and size \& shape of supercell was optimized. Using the optimized structure we generated 47 different supercells by removing one nitrogen at time. Each was fully relaxed using the MTP potential, and the structure with the lowest energy was used for generation of structure with one less N atom. This process was repeated iteratively until all 48 N atoms were removed. For each stoichiometry, structures generated via both routes were compared, and the one with the lower formation energy was identified as the most stable configuration.
\vspace{1\baselineskip}
\\(ii)For N/Ti ratios between 0.5 and 1.0, intermediate structures corresponding to Ti\textsubscript{2}N ,Ti\textsubscript{3}N\textsubscript{2}, Ti\textsubscript{4}N\textsubscript{3}, Ti\textsubscript{6}N\textsubscript{5} and  TiN compositions were generated. Nitrogen insertion was performed by identifying and occupying unfilled octahedral interstitial sites formed by surrounding Ti atoms. After each insertion step, the structure was relaxed using the {MTP$_{0.75{-}0.25{-}0.25}$} potential, and the formation energy per atom was calculated. Among the generated configurations, the one with the lowest formation energy was selected to proceed to the next insertion step. This insertion process was applied for  Ti\textsubscript{2}N → Ti\textsubscript{3}N\textsubscript{2},Ti\textsubscript{3}N\textsubscript{2} → Ti\textsubscript{4}N\textsubscript{3}, Ti\textsubscript{4}N\textsubscript{3} → Ti\textsubscript{6}N\textsubscript{5}, and Ti\textsubscript{6}N\textsubscript{5} → TiN transitions. Nitrogen removal followed the same systematic procedure described for the Ti\textsubscript{2}N → Ti transition and was sequentially applied for Ti\textsubscript{3}N\textsubscript{2}→ Ti\textsubscript{2}N, Ti\textsubscript{4}N\textsubscript{3} → Ti\textsubscript{3}N\textsubscript{2}, Ti\textsubscript{6}N\textsubscript{5} → Ti\textsubscript{4}N\textsubscript{3},and TiN → Ti\textsubscript{6}N\textsubscript{5}.
\vspace{1\baselineskip}

\begin{figure}[H]
  \centering
  \includegraphics[width=0.9\textwidth]{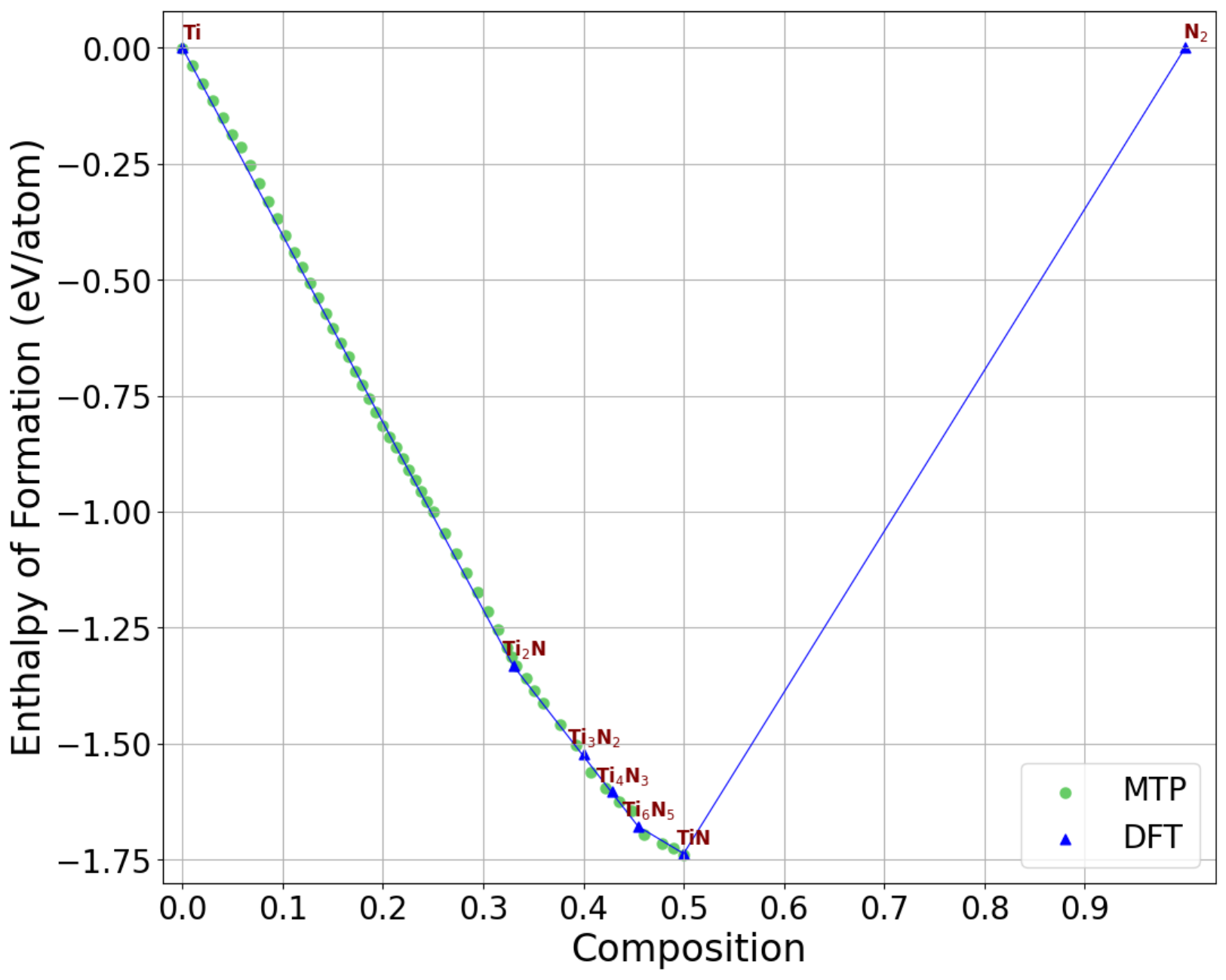}
  \caption{\textbf{Convex hull illustrating the stability of intermediate structures generated through insertion and removal of nitrogen atoms. Each intermediate points in the hull plot corresponds to specific composition derived from the most stable phase either by insertion or removal of Nitrogen.}}
  \label{fig:N_insertion_deletion}
\end{figure}
\vspace{1\baselineskip}
\noindent{(iii) For the Ti\textsubscript{2}N → Ti\textsubscript{3}N\textsubscript{2} transition, a tetragonal Ti\textsubscript{2}N supercell (Ti\textsubscript{96}N\textsubscript{48}) was used. Unoccupied octahedral voids which were formed by Ti were identified and sixteen nitrogen atoms were inserted into available octahedral voids to obtain Ti\textsubscript{64}N\textsubscript{32} (Ti\textsubscript{3}N\textsubscript{2}). For the Ti\textsubscript{3}N\textsubscript{2}→ Ti\textsubscript{2}N transition, Ti\textsubscript{96}N\textsubscript{64} with orthorhombic symmetry was used as the starting structure. A single nitrogen atom was randomly removed to generate Ti\textsubscript{96}N\textsubscript{63}, followed by systematic removal from all distinct sites, generating 63 configurations, as described for the Ti\textsubscript{2}N → Ti transition. Each was relaxed, and the configuration with the lowest formation energy was chosen for the next iteration, continuing until Ti\textsubscript{96}N\textsubscript{48} was reached.}
\vspace{1\baselineskip}
\\(iv)In the Ti\textsubscript{3}N\textsubscript{2} → Ti\textsubscript{4}N\textsubscript{3} transition, the orthorhombic Ti\textsubscript{96}N\textsubscript{64} structure served as the initial structure, and eight nitrogen atoms were inserted to produce Ti\textsubscript{96}N\textsubscript{72}. The Ti\textsubscript{4}N\textsubscript{3} → Ti\textsubscript{3}N\textsubscript{2}, involved plucking eight nitrogen atoms from the Ti\textsubscript{96}N\textsubscript{72} monoclinic structure, following the same energy-based selection protocol.
\vspace{1\baselineskip}
\\(v) For Ti\textsubscript{4}N\textsubscript{3} → Ti\textsubscript{6}N\textsubscript{5}, nitrogen insertion into Ti\textsubscript{96}N\textsubscript{72} (monoclinic) was carried out by adding eight nitrogen atoms, resulting in Ti\textsubscript{96}N\textsubscript{80}. The reverse, Ti\textsubscript{6}N\textsubscript{5} → Ti\textsubscript{4}N\textsubscript{3}, was performed by removing eight nitrogen atoms from the monoclinic Ti\textsubscript{96}N\textsubscript{80} structure.
\vspace{1\baselineskip}
\\(vi) Finally, the Ti\textsubscript{6}N\textsubscript{5} → TiN transition was modeled by inserting sixteen nitrogen atoms into monoclinic Ti\textsubscript{96}N\textsubscript{80}, forming Ti\textsubscript{96}N\textsubscript{96} (TiN). For the TiN → Ti\textsubscript{6}N\textsubscript{5} transition, sixteen nitrogen atoms were systematically removed from Ti\textsubscript{96}N\textsubscript{96}, again selecting the lowest-energy configuration at each step.

The details of the creation of diverse structures through N-atom insertion and deletion is as shown in Figure~\ref{fig:N_insertion_deletion}. The minimum formation energy from either of the process is considered, selecting the more energetically favourable structure.

\section{Conclusion}
We developed a MTP based interatomic potential for Ti-N system. Although various known structures of Ti-N system like  Ti\textsubscript{2}N, Ti\textsubscript{3}N\textsubscript{2}, Ti\textsubscript{6}N\textsubscript{5}, Ti\textsubscript{4}N\textsubscript{3} alongwith Ti , TiN and solid solutions of N in Ti belongs to different space group. There is structural similarity between Ti and solid solutions. Also Ti\textsubscript{3}N\textsubscript{2}, Ti\textsubscript{6}N\textsubscript{5}, Ti\textsubscript{4}N\textsubscript{3} are ordered N vacancy in TiN. This information helps us choose diverse training dataset. Inclusions of forces and stresses other than energy was essential to improve the energy prediction specifically for the systems that were not part of the training dataset. RMSE error in prediction of formation energy of systems that was part of training dataset is 2.1 meV/atom and for system that was not part of the training dataset error is 6.8 meV/atom. The KDE plot was utilized for visualization of the distribution of formation energy errors as compared to DFT reference values and the energy  peaks were found to have a maximum value of 3.8 meV/atom for the training dataset and 7.6 meV/atom for the systems that are not part of the training dataset. Elastic constants were also accurately predicted for systems that was not part of the training dataset. Molecular Dynamics for all the Ti-N system under NPT condition till room temperature exhibited total energy fluctuations within (1-10)\% indicating the thermodynamic stability. This behaviour is valid for all the Ti-N system, whether it is a part of the training dataset or not. We also show that structures with N/Ti ratios ranging from 0 to 1 can be thermodynamically stable. A maximum deviation of 10 meV/atom from the convex hull plot of formation energy at 0K was observed for predicted structures, this indicates the structures can be thermodynamic stability.

\bibliographystyle{elsarticle-num} 
\bibliography{sources}

\end{document}


\maketitle

\section*{Elastic Constants via Energy–Strain Method}
The given document presents the different strain modes and their corresponding elastic energy expressions used to extract second-order elastic constants for various Ti-N compounds. This approach is based on symmetry resolved strain energy relationships following Voigt notation. The general energy-strain expression is:

\begin{equation}
\frac{\Delta E}{V} = \frac{1}{2} \sum_{i,j} C_{ij} \varepsilon_i \varepsilon_j
\end{equation}

Each compound's crystal system and corresponding space group determine the number and type of strain modes used. The tables below list the deformation modes and energy expressions per system.

\section{Crystal Systems and Strain Modes}

\subsection{Monoclinic Systems}
Table~\ref{tab:monoclinic} lists the 13 independent strain modes used for monoclinic crystals. These include uniaxial, shear, and mixed deformation modes.

\begin{table}[h!]
\centering
\caption{Strain modes and derived elastic constants for monoclinic systems.}
\label{tab:monoclinic}
\begin{tabular}{ccc}
\toprule
\textbf{Strain Index} & \textbf{Strain Vector $\boldsymbol{\varepsilon}$} & \textbf{Elastic Energy $\frac{\Delta E}{V}$} \\
\midrule
1 & $(\delta, 0, 0, 0, 0, 0)$ & $\frac{1}{2} C_{11} \delta^2$ \\
2 & $(0, \delta, 0, 0, 0, 0)$ & $\frac{1}{2} C_{22} \delta^2$ \\
3 & $(0, 0, \delta, 0, 0, 0)$ & $\frac{1}{2} C_{33} \delta^2$ \\
4 & $(0, 0, 0, \delta, 0, 0)$ & $\frac{1}{2} C_{44} \delta^2$ \\
5 & $(0, 0, 0, 0, \delta, 0)$ & $\frac{1}{2} C_{55} \delta^2$ \\
6 & $(0, 0, 0, 0, 0, \delta)$ & $\frac{1}{2} C_{66} \delta^2$ \\
7 & $(\delta, \delta, 0, 0, 0, 0)$ & $\frac{1}{2}(C_{11} + C_{12} + 2C_{12}) \delta^2$ \\
8 & $(\delta, 0, \delta, 0, 0, 0)$ & $\frac{1}{2}(C_{11} + C_{33} + 2C_{13}) \delta^2$ \\
9 & $(0, \delta, \delta, 0, 0, 0)$ & $\frac{1}{2}(C_{22} + C_{33} + 2C_{23}) \delta^2$ \\
10 & $(\delta, 0, 0, \delta, 0, 0)$ & $\frac{1}{2}(C_{11} + C_{44} + 2C_{14}) \delta^2$ \\
11 & $(0, \delta, 0, 0, \delta, 0)$ & $\frac{1}{2}(C_{22} + C_{55} + 2C_{25}) \delta^2$ \\
12 & $(0, 0, \delta, 0, 0, \delta)$ & $\frac{1}{2}(C_{33} + C_{66} + 2C_{36}) \delta^2$ \\
13 & $(0, 0, 0, \delta, \delta, 0)$ & $\frac{1}{2}(C_{44} + C_{55} + 2C_{45}) \delta^2$ \\
\bottomrule
\end{tabular}
\end{table}

\subsection{Hexagonal Systems}
Table~\ref{tab:hexagonal} provides strain modes used for hexagonal systems.

\begin{table}[h!]
\centering
\caption{Strain modes and elastic energy for hexagonal systems.}
\label{tab:hexagonal}
\begin{tabular}{ccc}
\toprule
\textbf{Strain Index} & \textbf{Strain Vector $\boldsymbol{\varepsilon}$} & \textbf{Elastic Energy $\frac{\Delta E}{V}$} \\
\midrule
1 & $(\delta, 0, 0, 0, 0, 0)$ & $\frac{1}{2} C_{11} \delta^2$ \\
2 & $(0, \delta, 0, 0, 0, 0)$ & $\frac{1}{2} C_{11} \delta^2$ \\
3 & $(0, 0, \delta, 0, 0, 0)$ & $\frac{1}{2} C_{33} \delta^2$ \\
4 & $(\delta, \delta, 0, 0, 0, 0)$ & $\frac{1}{2}(C_{11} + C_{12} + 2C_{12}) \delta^2$ \\
5 & $(0, 0, \delta, 0, 0, \delta)$ & $\frac{1}{2}(C_{33} + C_{13} + 2C_{13}) \delta^2$ \\
6 & $(0, 0, 0, 0, 0, \delta)$ & $\frac{1}{2} C_{44} \delta^2$ \\
\bottomrule
\end{tabular}
\end{table}

\subsection{Cubic Systems}
Table~\ref{tab:cubic} shows typical strain modes for cubic crystals.

\begin{table}[h!]
\centering
\caption{Strain modes and elastic energy expressions for cubic systems.}
\label{tab:cubic}
\begin{tabular}{ccc}
\toprule
\textbf{Strain Index} & \textbf{Strain Vector $\boldsymbol{\varepsilon}$} & \textbf{Elastic Energy $\frac{\Delta E}{V}$} \\
\midrule
1 & $(\delta, 0, 0, 0, 0, 0)$ & $\frac{1}{2} C_{11} \delta^2$ \\
2 & $(\delta, \delta, \delta, 0, 0, 0)$ & $\frac{1}{2}(C_{11} + 2C_{12}) \delta^2$ \\
3 & $(0, 0, 0, 0, 0, \delta)$ & $\frac{1}{2} C_{44} \delta^2$ \\
\bottomrule
\end{tabular}
\end{table}

\subsection{Orthorhombic system}
Table~\ref{tab:orthorhombic} shows typical strain modes for orthorhombic crystals
\begin{table}[h!]
\centering
\caption{Strain modes and elastic energy expressions for orthorhombic systems.}
\label{tab:orthorhombic}
\begin{tabular}{ccc}
\toprule
\textbf{Strain index} & \textbf{Strain vector $\varepsilon$} & \textbf{Elastic energy $\frac{\Delta E}{V}$} \\
\midrule
1 & ($\delta$, 0, 0, 0, 0, 0) & $\frac{1}{2} C_{11} \delta^2$ \\
2 & (0, $\delta$, 0, 0, 0, 0) & $\frac{1}{2} C_{22} \delta^2$ \\
3 & (0, 0, $\delta$, 0, 0, 0) & $\frac{1}{2} C_{33} \delta^2$ \\
4 & (0, 0, 0, $\delta$, 0, 0) & $\frac{1}{2} C_{44} \delta^2$ \\
5 & (0, 0, 0, 0, $\delta$, 0) & $\frac{1}{2} C_{55} \delta^2$ \\
6 & (0, 0, 0, 0, 0, $\delta$) & $\frac{1}{2} C_{66} \delta^2$ \\
7 & ($\delta$, $\delta$, 0, 0, 0, 0) & $\frac{1}{2} (C_{11} + C_{22} + 2C_{12}) \delta^2$ \\
8 & ($\delta$, 0, $\delta$, 0, 0, 0) & $\frac{1}{2} (C_{11} + C_{33} + 2C_{13}) \delta^2$ \\
9 & (0, $\delta$, $\delta$, 0, 0, 0) & $\frac{1}{2} (C_{22} + C_{33} + 2C_{23}) \delta^2$ \\
\bottomrule
\end{tabular}
\end{table}
\subsection{Tetragonal system}
Table~\ref{tab:tetragonal} shows typical strain modes for tetragonal crystals.
\begin{table}[h!]
\centering
\caption{Strain modes and elastic energy expressions for tetragonal systems.}
\label{tab:tetragonal}
\begin{tabular}{ccc}
\toprule
\textbf{Strain index} & \textbf{Strain vector $\varepsilon$} & \textbf{Elastic energy $\frac{\Delta E}{V}$} \\
\midrule
1 & ($\delta$, 0, 0, 0, 0, 0) & $\frac{1}{2} C_{11} \delta^2$ \\
2 & (0, $\delta$, 0, 0, 0, 0) & $\frac{1}{2} C_{22} \delta^2$ \\
3 & (0, 0, $\delta$, 0, 0, 0) & $\frac{1}{2} C_{33} \delta^2$ \\
4 & (0, 0, 0, $\delta$, 0, 0) & $\frac{1}{2} C_{44} \delta^2$ \\
5 & (0, 0, 0, 0, $\delta$, 0) & $\frac{1}{2} C_{55} \delta^2$ \\
6 & (0, 0, 0, 0, 0, $\delta$) & $\frac{1}{2} C_{66} \delta^2$ \\
7 & ($\delta$, $\delta$, 0, 0, 0, 0) & $\frac{1}{2} (C_{11} + C_{22} + 2C_{12}) \delta^2$ \\
8 & ($\delta$, 0, $\delta$, 0, 0, 0) & $\frac{1}{2} (C_{11} + C_{33} + 2C_{13}) \delta^2$ \\
9 & (0, $\delta$, $\delta$, 0, 0, 0) & $\frac{1}{2} (C_{22} + C_{33} + 2C_{23}) \delta^2$ \\
\bottomrule
\end{tabular}
\end{table}

\begin{table}[h!]
\centering
\caption{Mapping of compounds to crystal systems and strain tables.}
\label{tab:compound_map}
\begin{tabular}{lll}
\toprule
\textbf{Compound} & \textbf{Crystal System (Space Group)} & \textbf{Strain Table} \\
\midrule
A–B & Monoclinic (SG: P2$_1$/m) & Table~\ref{tab:monoclinic} \\
X–Y & Hexagonal (SG: P6$_3$/mmc) & Table~\ref{tab:hexagonal} \\
M–N & Cubic (SG: Fm$\bar{3}$m) & Table~\ref{tab:cubic} \\
\bottomrule
\end{tabular}
\end{table}